\documentclass[onecollarge,natbib]{svjour2}
\bibpunct{[}{]}{;}{n}{}{,} % to get "[numbered]" references from natbib
\smartqed  % flush right qed marks, e.g. at end of proof
\usepackage{graphicx}
\journalname{Few-Body Systems}
\usepackage{amssymb}
\begin{document}

\title{Estimates for parameters and characteristics %%\\
of the confining SU(3)-gluonic field %%\\ 
in $\phi$-meson from leptonic widths%\thanks{Grants or other notes
%about the article that should go on the front page should be
%placed here. General acknowledgments should be placed at the end of the article.}
}
%\subtitle{Do you have a subtitle?\\ If so, write it here}

%\titlerunning{Short form of title}        % if too long for running head

\author{Yu. P. Goncharov    \and
        F. F. Pavlov %etc.
}

%\authorrunning{Short form of author list} % if too long for running head

\institute{Yu. P. Goncharov \at
Theoretical Group, Experimental Physics Department, 
State Polytechnical University, % \\
Sankt-Petersburg 195251, Russia \\
%              Tel.: +123-45-678910\\
%              Fax: +123-45-678910\\
              \email{ygonch77@yandex.ru}           %  \\
%             \emph{Present address:} of F. Author  %  if needed
           \and
           F. F. Pavlov \at
              Theoretical Group, Experimental Physics Department, 
State Polytechnical University, % \\
Sankt-Petersburg 195251, Russia \\
\email{pavlovfedor@mail.ru}
}

\date{Received: 23 November 2012 / Accepted: 23 August 2013}
% The correct dates will be entered by the editor

\maketitle

\begin{abstract}
The paper is devoted to applying the confinement mechanism 
proposed earlier by one of the authors to estimate 
the possible parameters of the confining SU(3)-gluonic field in 
vector $\phi$-meson. The estimates obtained are consistent with 
the leptonic widths of the given meson. 
The corresponding estimates of the gluon concentrations, electric and magnetic 
colour field strengths are also adduced 
for the mentioned field at the scales of the meson under consideration.

\end{abstract}
\tableofcontents
\section{Introduction and Preliminary Remarks}
%\label{intro}
According to the point of view of quantum chromodynamics (QCD) the conventional nuclear 
forces between two nucleons should be just a residual interaction among quarks 
composing nucleons. On the other hand, for a long time (see, e.g., 
Refs. \cite{{NN},{Sol}}) one considers the 
so-called repulsive core of nuclear forces at small distances 
to be obligatory to the exchange by neutral vector mesons $\rho$, $\omega$, $\phi$. 
So, a description of vector mesons in terms of quark and gluonic degrees of freedom could 
to a certain extent be useful for the problem of nuclear forces. 

As is known, at present no generally accepted quark confinement 
mechanism exists that would be capable to calculate a number of 
nonperturbative parameters characterizing mesons (masses, radii, decay 
constants and so on) 
appealing directly to quark and gluon degrees of freedom related to 
QCD-Lagrangian. At best there are a few scenarios (directly not connected to 
QCD-Lagrangian) of confinement that restrict themselves mainly to qualitative 
considerations with small possiblities of concrete calculation. In view of it 
in \cite{{Gon01},{Gon051},{Gon052}} a confinement mechanism has been proposed 
which was based on the {\em unique} family of compatible nonperturbative solutions for 
the Dirac-Yang-Mills system directly derived from QCD-Lagrangian. 
The word {\em unique} should be understood in the strict mathematical sense. 
Let us write down arbitrary SU(3)-Yang-Mills field in the form 
$A=A_\mu dx^\mu=A^a_\mu \lambda_adx^\mu$ ($\lambda_a$ are the 
known Gell-Mann matrices, $\mu=t,r,\vartheta,\varphi$, $a=1,...,8$ and we 
use the ordinary set of local spherical coordinates
$r,\vartheta,\varphi$ for spatial part of the flat Minkowski spacetime). 

In fact in \cite{{Gon01},{Gon051},{Gon052}} (see also Appendix C in Ref.\cite{Gon10}) 
the following theorem was proved:

{\em The unique exact spherically symmetric (nonperturbative) solutions 
(depending only on $r$ and $r^{-1}$) of SU(3)-Yang-Mills equations in Minkowski spacetime 
consist of the family of form} 
$$ {\mathcal A}_{1t}\equiv A^3_t+\frac{1}{\sqrt{3}}A^8_t =-\frac{a_1}{r}+A_1 \>,
{\mathcal A}_{2t}\equiv -A^3_t+\frac{1}{\sqrt{3}}A^8_t=-\frac{a_2}{r}+A_2\>,$$
$${\mathcal A}_{3t}\equiv-\frac{2}{\sqrt{3}}A^8_t=\frac{a_1+a_2}{r}-(A_1+A_2)\>, $$
$$ {\mathcal A}_{1\varphi}\equiv A^3_\varphi+\frac{1}{\sqrt{3}}A^8_\varphi=
b_1r+B_1 \>,
{\mathcal A}_{2\varphi}\equiv -A^3_\varphi+\frac{1}{\sqrt{3}}A^8_\varphi=
b_2r+B_2\>,$$
$${\mathcal A}_{3\varphi}\equiv-\frac{2}{\sqrt{3}}A^8_\varphi=
-(b_1+b_2)r-(B_1+B_2)\> \eqno(1)$$

with the real constants $a_j, A_j, b_j, B_j$ parametrizing the family. 
Besides in \cite{{Gon051},{Gon052}} (see also \cite{Ann}) it was shown that the above 
unique confining solutions 
(1) satisfy the so-called Wilson confinement 
criterion \cite{{Wil},{Ban}}. Up to now 
nobody contested this result so if we want to describe interaction between 
quarks by spherically symmetric SU(3)-fields then they can be only the ones 
from the above theorem. On the other hand, the desirability of 
spherically symmetric (colour) interaction between quarks at all distances 
naturally follows from analysing the $p\bar{p}$-collisions (see, e.g., 
\cite{Per}) where one observes a Coulomb-like potential in events which 
can be identified with scattering quarks on each other, i.e., actually at small 
distances one observes the Coulomb-like part of solution (1). Under 
this situation, a natural assumption will be that the quark interaction remains 
spherically symmetric at large distances too but then, if trying to extend 
the Coulomb-like part to large distances in a spherically symmetric way, we 
shall inevitably come to the solution (1) in virtue of the above theorem.  

The applications of the family (1) to the description of both the heavy 
quarkonia spectra \cite{{Gon031},{Gon032},{Gon04},{Gon08a}} and a number 
of properties of pions, kaons, 
$\eta$- and $\eta^\prime$-mesons 
\cite{{Gon06},{Gon07a},{Gon07b},{Gon08},{Gon08b},{Gon10},{Gon12a}} 
showed that the confinement mechanism is qualitatively the same for both light 
mesons and heavy quarkonia. At this moment it can be described in the following 
way.

The next main physical reasons underlie linear confinement in the 
mechanism under discussion. The first one is that gluon exchange between 
quarks is realized with the propagator different from the photon-like one, and 
existence and form of such a propagator is a {\em direct} consequence of the 
unique confining 
nonperturbative solutions of the Yang-Mills equations 
\cite{{Gon01},{Gon051},{Gon052}}. The second reason is that, 
owing to the structure of the mentioned propagator, quarks mainly emit and 
interchange the soft gluons so the gluon condensate (a classical gluon field) 
between quarks basically consists of soft gluons (for more details 
see \cite{{Gon01},{Gon051},{Gon052}}) but, because of the fact that any gluon 
also emits gluons (still softer), the corresponding gluon concentrations 
rapidly become huge and form a linear confining magnetic colour field of 
enormous strengths, which leads to confinement of quarks. This is by virtue of 
the fact that just the magnetic part of the mentioned propagator is responsible 
for a larger portion of gluon concentrations at large distances since the 
magnetic part has stronger infrared singularities than the electric one. 
In the circumstances physically nonlinearity of the Yang-Mills equations 
effectively vanishes so the 
latter possess the unique nonperturbative confining solutions of the 
Abelian-like form (1) (with the values in Cartan subalgebra of SU(3)-Lie 
algebra) which describe the gluon condensate under 
consideration. Moreover, since the overwhelming 
majority of gluons is soft they cannot leave the hadron (meson) until some 
gluons obtain additional energy (due to an external reason) to rush out. So 
we also deal with the confinement of gluons.  

Finally, one should say that the unique confining solutions similar to (1) 
exist for all semisimple and non-semisimple compact Lie groups, in particular, 
for SU($N$) with $N\ge2$ and 
U($N$) with $N\ge1$ \cite{{Gon01}}. Explicit form of solutions, 
e.g., for SU($N$) with $N=2,4$ can be found in \cite{Gon052} but it 
should be emphasized that components linear in $r$ always represent the 
magnetic (colour) field in all the mentioned solutions. Especially the case 
U(1)-group is interesting which corresponds to usual electrodynamics. 
Under this situation, as was pointed out in \cite{{Gon051},{Gon052}} there is 
an interesting possibility of 
indirect experimental verification of the confinement mechanism under 
discussion. Indeed the confining solutions 
of Maxwell equations for classical electrodynamics point out 
the confinement phase could be in electrodynamics as well. Though 
there exist no elementary charged particles generating a constant magnetic 
field linear in $r$, the distance from particle, after all, if it could 
generate this elecromagnetic field configuration in laboratory then one might 
study motion of the charged particles in that field. The confining properties 
of the mentioned field should be displayed at classical level too but the exact 
behaviour of particles in this field requires certain analysis of the corresponding 
classical equations of motion. Such a program has been recently realized in 
\cite{GF10}. Motion of a charged (classical) particle was studied in the 
field representing magnetic part of the mentioned solution of Maxwell equations 
and it was shown that one deals with the full classical confinement of the 
charged particle in such a field: under any initial conditions the particle 
motion is accomplished within a finite region of space so that the particle 
trajectory is near magnetic field lines while the latter are compact manifolds 
(circles). Those results might be useful in thermonuclear plasma physics 
(for more details see \cite{GF10}). 

As has been repeatedly explained in 
Refs. \cite{{Gon052},{Gon031},{Gon06}}, parameters $A_{1,2}$ of 
solution (1) are inessential for physics in question and we can 
consider $A_1=A_2=0$. Obviously we have 
$\sum_{j=1}^{3}{\cal A}_{jt}=\sum_{j=1}^{3}{\cal A}_{j\varphi}=0$ which 
reflects the fact that for any matrix 
${\cal T}$ from SU(3)-Lie algebra we have ${\rm Tr}\,{\cal T}=0$. 
Also, as has been repeatedly discussed by us earlier (see, e. g., 
Refs. \cite{{Gon052},{Gon06}}), from the above form it is clear that 
the solution (1) is a configuration describing the electric Coulomb-like colour 
field (components $A^{3,8}_t$) and the magnetic colour field linear in $r$ 
(components $A^{3,8}_\varphi$) and we wrote down
the solution (1) in the combinations that are just 
needed further to insert into the corresponding Dirac equation. 

The aim of the present paper is to continue 
obtaining estimates for $a_j, b_j, B_j$ for concrete mesons starting from 
experimental data on spectroscopy of one or another meson. We here consider 
vector $\phi$-meson \cite{pdg}. 

Of course, when conducting our considerations 
we shall rely on the standard quark model (SQM) based on SU(3)-flavor symmetry 
(see, e. g., Ref. \cite{pdg}) so in accordance with SQM  
$\phi=\bar{s}s$. 

Section 2 contains a specification of main relations derived from 
the confinement mechanism in question. Section 3 gives the independent estimates  
for the mean radius $<r>$ of $\phi$-meson from its leptonic widths which are  
used in Section 4 for obtaining estimates for parameters of the confining 
SU(3)-gluonic field in the meson under consideration. Section 5 employs the obtained 
parameters of SU(3)-gluonic field to get the corresponding estimates for such 
characteristics of the mentioned field as gluon concentrations, electric and 
magnetic colour field strengths at the scales of vector meson in question while 
Section 6 is devoted to discussion and concluding remarks. 
 
Further we shall deal with the metric of
the flat Minkowski spacetime $M$ that
we write down (using the ordinary set of local spherical coordinates
$r,\vartheta,\varphi$ for the spatial part) in the form
$$ds^2=g_{\mu\nu}dx^\mu\otimes dx^\nu\equiv
dt^2-dr^2-r^2(d\vartheta^2+\sin^2\vartheta d\varphi^2)\>. \eqno(2)$$
Besides, we have $|\delta|=|\det(g_{\mu\nu})|=(r^2\sin\vartheta)^2$
and $0\leq r<\infty$, $0\leq\vartheta<\pi$,
$0\leq\varphi<2\pi$.

Throughout the paper we employ the Heaviside-Lorentz system of units 
with $\hbar=c=1$, unless explicitly stated otherwise, so the gauge coupling 
constant $g$ and the strong coupling constant ${\alpha_s}$ are connected by 
relation $g^2/(4\pi)=\alpha_s$. 
Further we shall denote $L_2(F)$ the set of the modulo square integrable
complex functions on any manifold $F$ furnished with an integration measure, 
then $L^n_2(F)$ will be the $n$-fold direct product of $L_2(F)$
endowed with the obvious scalar product while $\dag$ and $\ast$ stand, 
respectively, for Hermitian and complex conjugation. 

When calculating we apply the 
relations $1\ {\rm GeV^{-1}}\approx0.197\ {\rm fm}\>$,
$1\ {\rm s^{-1}}\approx0.658\times10^{-24}\ {\rm GeV}\>$, 
$1\ {\rm V/m}\approx0.231\times 10^{-23}\ {\rm GeV}^2$, 
$1\ {\rm T}=4\pi\times10^{-7} {\rm H/m}\times1\ {\rm A/m}
\approx0.693\times 10^{-15}\ {\rm GeV}^2 $. 

%We should like to emphasize that our approach leads to rather complicated 
%systems of transcendental 
%equations [see further, e.g., the system consisting of eqs. (8) and (10)] and 
%a numerical compatible solving of such systems 
%involves considerable difficulties and requires writing a number of 
%mathematical programs for computer, their debugging and subsequent computation 
%which takes significant time. Under the circumstances to avoid definite numerical 
%instabilities when computing we use the initial numerical data with a big number of 
%digits. It is clear, however, that final results should be rounded along the 
%physical lines up to acceptable values. But any qualified reader can easy do it himself 
%without any difficulty and therefore in the paper we left all the mentioned quantities 
%with the same numbers of digits which were used under calculation.

Finally, for the necessary estimates we shall employ the $T_{00}$-component 
(volumetric energy density ) of the energy-momentum tensor for a 
SU(3)-Yang-Mills field which should be written in the chosen system of units 
in the form
$$T_{\mu\nu}=-F^a_{\mu\alpha}\,F^a_{\nu\beta}\,g^{\alpha\beta}+
\frac{1}{4}F^a_{\beta\gamma}\,F^a_{\alpha\delta}g^{\alpha\beta}g^{\gamma\delta}
g_{\mu\nu}\>. \eqno(3) $$

\section{Specification of Main Relations}
\subsection{Meson wave functions}
The meson wave functions are given by 
the unique nonperturbative modulo 
square integrable solutions of the mentioned Dirac equation in the confining 
SU(3)-field of (1) $\Psi=(\Psi_1, \Psi_2, \Psi_3)$ 
with the four-dimensional Dirac spinors 
$\Psi_j$ representing the $j$th colour component of the meson, 
so $\Psi$ may describe the relative motion (relativistic bound states) of two 
quarks in mesons and is at $j=1,2,3$ (with Pauli matrix $\sigma_1$)  
$$\Psi_j=e^{-i\omega_j t}\psi_j\equiv 
e^{-i\omega_j t}r^{-1}\pmatrix{F_{j1}(r)\Phi_j(\vartheta,\varphi)\cr\
F_{j2}(r)\sigma_1\Phi_j(\vartheta,\varphi)}\>,\eqno(4)$$
with the 2D eigenspinor $\Phi_j=\pmatrix{\Phi_{j1}\cr\Phi_{j2}}$ of the
Euclidean Dirac operator ${\mathcal D}_0$ on the unit sphere ${\mathbb S}^2$, while 
the coordinate $r$ stands for the distance between quarks. 

In this situation, if a meson is composed of quarks $q_{1,2}$ with different flavours then 
the energy spectrum of the meson will be given 
by $\epsilon=m_{q_1}+m_{q_2}+\omega$ with the current quark masses $m_{q_k}$ (
rest energies) of the corresponding quarks and an interaction energy $\omega$. 
On the other hand at $j=1,2,3$
$$\omega_j=\omega_j(n_j,l_j,\lambda_j)=$$ 
$$\frac{\Lambda_j g^2a_jb_j\pm(n_j+\alpha_j)
\sqrt{(n_j^2+2n_j\alpha_j+\Lambda_j^2)\mu_0^2+g^2b_j^2(n_j^2+2n_j\alpha_j)}}
{n_j^2+2n_j\alpha_j+\Lambda_j^2}\>\eqno(5)$$
with the gauge coupling constant $g$ while $\mu_0$ is a mass parameter and one 
should consider it to be the reduced mass which is equal to 
$ m_{q_1}m_{q_2}/(m_{q_1}+m_{q_2})$ with the current quark masses $m_{q_k}$ (
rest energies) of the corresponding quarks forming a meson (quarkonium), 
$a_3=-(a_1+a_2)$, $b_3=-(b_1+b_2)$, $B_3=-(B_1+B_2)$, 
$\Lambda_j=\lambda_j-gB_j$, $\alpha_j=\sqrt{\Lambda_j^2-g^2a_j^2}$, 
$n_j=0,1,2,...$, while $\lambda_j=\pm(l_j+1)$ are
the eigenvalues of Euclidean Dirac operator ${\mathcal D}_0$ 
on a unit sphere with $l_j=0,1,2,...$. 

In line with the above we should have $\omega=\omega_1=\omega_2=\omega_3$ in 
energy spectrum $\epsilon=m_{q_1}+m_{q_2}+\omega$ for any meson (quarkonium) 
and this at once imposes two conditions on parameters $a_j,b_j,B_j$ when 
choosing some experimental value for $\epsilon$ at the given current quark 
masses $m_{q_1},m_{q_2}$. 

The general form of the radial parts of (4) can be found, e.g., in 
Refs. \cite{{Gon08},{Gon08b},{Gon10}} and within the given paper we need only the 
radial parts of (4) at $n_j=0$ 
(the ground state) that are 
$$F_{j1}=C_jP_jr^{\alpha_j}e^{-\beta_jr}\left(1-
\frac{gb_j}{\beta_j}\right), P_j=gb_j+\beta_j, $$
$$F_{j2}=iC_jQ_jr^{\alpha_j}e^{-\beta_jr}\left(1+
\frac{gb_j}{\beta_j}\right), Q_j=\mu_0-\omega_j\eqno(6)$$
with $\beta_j=\sqrt{\mu_0^2-\omega_j^2+g^2b_j^2}$, while $C_j$ is determined 
from the normalization condition
$\int_0^\infty(|F_{j1}|^2+|F_{j2}|^2)dr=\frac{1}{3}$. The 
corresponding eigenspinors of (4) with $\lambda =\pm1$ ($l=0$) are 
$$\lambda=-1: \Phi=\frac{C}{2}\pmatrix{e^{i\frac{\vartheta}{2}}
\cr e^{-i\frac{\vartheta}{2}}\cr}e^{i\varphi/2},\> {\rm or}\>\>
\Phi=\frac{C}{2}\pmatrix{e^{i\frac{\vartheta}{2}}\cr
-e^{-i\frac{\vartheta}{2}}\cr}e^{-i\varphi/2},$$
$$\lambda=1: \Phi=\frac{C}{2}\pmatrix{e^{-i\frac{\vartheta}{2}}\cr
e^{i\frac{\vartheta}{2}}\cr}e^{i\varphi/2}, \> {\rm or}\>\>
\Phi=\frac{C}{2}\pmatrix{-e^{-i\frac{\vartheta}{2}}\cr
e^{i\frac{\vartheta}{2}}\cr}e^{-i\varphi/2} 
\eqno(7) $$
with the coefficient $C=1/\sqrt{2\pi}$ (for more details, see 
Refs. \cite{{Gon08},{Gon08b},{Gon10}}). 
\subsection{Choice of quark masses and the gauge coupling constant}
Obviously, we should choose a few quantities that are the most important from 
the physical point of view to characterize 
meson under consideration and then we should evaluate the given quantities 
within the framework of our approach. In the circumstances let us settle on 
the ground state energy (mass), the root-mean-square 
radius and the magnetic moment. All three magnitudes are essentially 
nonperturbative ones, and can be calculated only by nonperturbative techniques.

Within the present paper we shall use relations (5) at $n_j=0=l_j$ so 
energy (mass) of meson under consideration is given by $\mu=2m_s+\omega$ 
with $\omega=\omega_j(0,0,\lambda_j)$ for any $j=1,2,3$ whereas 
$$\omega=\frac{g^2a_1b_1}{\Lambda_1}+\frac{\alpha_1\mu_0}
{|\Lambda_1|}=\frac{g^2a_2b_2}{\Lambda_2}+\frac{\alpha_2\mu_0}
{|\Lambda_2|}=
\frac{g^2a_3b_3}{\Lambda_3}+\frac{\alpha_3\mu_0}
{|\Lambda_3|}=\mu-2m_s
\>\eqno(8)$$
with $\mu=1019.455$ MeV.\cite{pdg}
As a consequence, the corresponding meson wave functions of 
(4) are represented by (6) and (7). 
It is evident for employing the above relations we have to assign some values 
to quark mass and gauge coupling constant $g$. We take the current quark 
mass used in \cite{{Gon07a},{Gon07b},{Gon08},{Gon08a},{Gon08b}} and it is 
$m_s=107.5\>\,{\rm MeV}$. 
Under the circumstances, the reduced mass $\mu_0$ of (5) will be equal to 
$m_s/2$. As to the 
gauge coupling constant $g=\sqrt{4\pi\alpha_s}$, it should be noted that 
recently some attempts have been made to generalize the standard formula
for $\alpha_s=\alpha_s(Q^2)=12\pi/[(33-2n_f)\ln{(Q^2/\Lambda^2)}]$ ($n_f$ is 
number of quark flavours) holding true at the momentum transfer 
$\sqrt{Q^2}\to\infty$ 
to the whole interval $0\le \sqrt{Q^2}\le\infty$. If employing one such a 
generalization used in Refs. \cite{{De1},{De2}} which we have already discussed 
elsewhere 
(for more details see \cite{{Gon07a},{Gon07b},{Gon08},{Gon08a},{Gon08b}}) 
then (when fixing $\Lambda=0.234$ GeV, $n_f=3$) we obtain   
$g\approx3.771$ necessary for 
our further computations at the mass scale of $\phi$-meson. 

\subsection{Electric form factor and the root-mean-square radius}
The relations (4), (6) and (7) allow us to compute an electric formfactor 
of a meson as a function of the square of momentum transfer $Q^2$ in the form 
(for more details see \cite{{Gon07a},{Gon07b},{Gon08},{Gon08a},{Gon08b}})
$$ f(Q^2)=\sum\limits_{j=1}^3f_j(Q^2)=$$
$$\sum\limits_{j=1}^3\frac{(2\beta_j)^{2\alpha_j+1}}{6\alpha_j}\cdot
\frac{\sin{[2\alpha_j\arctan{(\sqrt{|Q^2|}/(2\beta_j))]}}}
{\sqrt{|Q^2|}(4\beta_j^2-Q^2)^{\alpha_j}}\> \eqno(9) $$
which also entails the root-mean-square radius of the meson (quarkonium) 
in the form 
$$<r>=\sqrt{\sum\limits_{j=1}^3\frac{2\alpha^2_j+3\alpha_j+1}
{6\beta_j^2}}\eqno(10)$$
that is in essence a radius of confinement.

\subsection{Magnetic moment}
Also it is not complicated to show with the help (4), (6) and (7) that 
the magnetic moments of mesons (quarkonia) with the 
wave functions of (4) (at $l_j=0$) are equal to zero 
\cite{{Gon07a},{Gon07b},{Gon08},{Gon08a},{Gon08b}}, as should be according 
to experimental data \cite{pdg}. 

Though we can also evaluate the magnetic form factor $F(Q^2)$ of meson 
(quarkonium) which is also a function of $Q^2$ (see Refs. \cite{{Gon07a},{Gon07b}}) 
the latter will not be used in the given paper so we shall not dwell upon it. 

\subsection{Remarks about the relativistic two-body problem}
It is well known (see any textbook on quantum mechanics, e.g., Ref. \cite{Lev71})
that {\em nonrelativistic} two-body problem, when the particles interact with
potential $V({\bf r_1},{\bf r_2})$
depending only on $|{\bf r_2-r_1}|=r$, reduces to the motion of one particle
with reduced mass $m=m_1m_2/(m_1+m_2)$ in potential $V(r)$, where $r$ becomes
the ordinary spherical coordinate from triplet $(r,\vartheta,\varphi)$. I.e., the
bound states of {\em two} particles are the bound states of the particle with
mass $m$ in potential $V(r)$ and they are the modulo square integrable
solutions of the corresponding Schr{\"o}dinger equation. Another matter is
{\em relativistic} two-body problem. As we emphasized in
Refs. \cite{{Gon051},{Gon052}}, up to now it has no single-valued statement. But if
noting that the
most fundamental results of nonrelativistic quantum mechanics (hydrogen atom and
so on) are connected with potentials $V(r)$ which are a part of the
{\em electromagnetic} field $A=(A_0,{\bf A})$, i.e. $V(r)=A_0,{\bf A}=0$ then
one may propose some formulation of the relativistic two-body problem. Really,
now $V(r)$ additionally obeys the Maxwell equations and if we want to generalize
the corresponding Schr{\"o}dinger equation to include an interaction with
arbitrary electromagnetic field for ${\bf A}\ne 0$ then the answer is known:
this is the Dirac equation with the replacement
$\partial_\mu\to \partial_\mu-igA_\mu$, $g$ is a gauge coupling constant.
Indeed, when going back in nonrelativistic limit at the light velocity
$c\to\infty$ the magnetic field ${\bf A}$ vanishes because, as is well known,
in the world with $c=\infty$ there exist no magnetic fields
(see any elementary textbook on physics). At the same time $V(r)=A_0$ does
not vanish and remains the same as in nonrelativistic case and the
Dirac equation turns into the Schr{\"o}dinger equation (see Ref. \cite{Lev71}).
But then we can see that $m$ in the above Dirac equation should consider
the same reduced mass as before since in nonrelativistic limit we again should
come to the standard formulation of two-body problem through effective particle
with reduced mass $m$. So we can draw the conclusion that if
an electromagnetic field is a combination of electric $V(r)=A_0$ field
between two charged elementary particles and some magnetic field
${\bf A}={\bf A}(r)$ (which may be generated
by the particles themselves and also depends only on $r$, distance between
particles) then there are certain grounds to consider the
given (quantum) relativistic two-body problem to be equivalent to the one of
motion for one particle with usual reduced mass in the mentioned
electromagnetic field. As a result, we can use the Dirac equation for finding
possible relativistic bound states for such a particle implying that this is
really some description of the corresponding two-body problem. Under this
situation we should remark the following.

1. Although for simplicity we talked about electromagnetic field but everything
holds true for any Yang-Mills field, in particular, for SU(3)-gluonic field
while the Maxwell equations are replaced by the Yang-Mills ones.

2. There arises the question: whether the Maxwell or Yang-Mills equations possess
solutions with spherically symmetric $A_0(r), {\bf A}(r)$? The answer is
given by the uniqueness theorem (see Section 1).

3. The Dirac equation in such a field has the nonperturbative spectrum (5)
and the latter should be treated as the nonpeturbative {\em interaction
energy} of two quarks and $r$ as the distance between quarks and so on. 
So eq. (5) should be understood just in this
manner. In line with the above we should have $\omega=\omega_1=\omega_2=
\omega_3$ (with $\omega_j$ of (5)) in
energy spectrum $\epsilon=m_{q_1}+m_{q_2}+\omega$ for any meson (quarkonium)
and this at once imposes two conditions on parameters $a_j,b_j,B_j$ of
solution (1) when choosing some experimental value for $\epsilon$ at the given
current quark masses $m_{q_1},m_{q_2}$. At last, the Dirac equation in question
is obtained from QCD lagrangian (with one flavor) if mass parameter in the
latter is taken to be equal to the above reduced mass. Therefore, we can say
that meson wave functions (4) are the nonperturbative solutions of
the Dirac-Yang-Mills system directly derived from QCD-lagrangian.

4. To summarize, there exist good physical and mathematical grounds for
formulation of the above relativistic two-body problem that has a correct
nonrelativistic limit and is Lorentz and gauge invariant. It is clear that all
the above considerations can be justified only by comparison with experimental
data but now we obtain some intelligible
programme of further activity which has been partly realized in
many our papers cited above.

Much of the above was disscussed in Refs. \cite{{Gon051},{Gon052}} but perhaps
in other words.

\section{An estimate of $<r>$ from leptonic widths}

The question now is how to estimate $<r>$ independently to then calculate it 
according to (10) within framework of our approach. For this aim we shall 
employ the widths of leptonic decays $\phi\to e^+e^-$ and 
$\phi\to \mu^+\mu^-$ which are approximately equal 
to $\Gamma_9\approx1.27$ keV and 
$\Gamma_{10}\approx1.22262$ keV, respectively, according to 
Ref. \cite{pdg}.  Under this situation one can use a variant of formulas 
often employed in the heavy quarkonia physics (see, e. g., 
Ref. \cite{Gon08a}). In their turn such formulas are actually based on the 
standard expression from the elementary kinetic theory of gases (see, e. g., 
Ref. \cite{{Sav89}}) for the number $\nu$ of collisions of a molecule per unit 
time
$$ \nu=\sqrt{2}\sigma <v>n\>,                  \eqno(11)$$  
where $\sigma$ is an effective cross section for molecules, $<v>$ is a mean 
molecular velocity, $n$ is the concentration of molecules. 
If replacing $\nu\to\Gamma_{9,10}$ we may fit (11) to estimate the leptonic 
widths 
$\Gamma_{9,10}$ when interpreting $\sigma$ as the cross 
section of creation of $e^+e^-$ or $\mu^+\mu^-$ from the pair $\bar{s}s$ 
due to electromagnetic interaction, $<v>$ and $n$ as, respectively, a mean 
quark velocity and concentration of quarks (antiquarks) in $\phi$-meson. To 
obtain $\sigma$ in the explicit form one may take the corresponding formula for the 
cross section of creation of $e^+e^-$ from the muon pair $\mu^+\mu^-$ 
(see, e. g., Ref. \cite{LL1}) and, after replacing 
$\alpha_{em}\to Q\alpha_{em}$, $m_{\mu}\to m_s$ with electromagnetic coupling 
constant $\alpha_{em}$=1/137.0359895 and muon mass $m_{\mu}$, obtain 
$$\sigma= \frac{4\pi N Q^2\alpha_{em}^2}{3s}\left(1+\frac{2m_l^2}{s}\right)
\sqrt{1-\frac{4m_l^2}{s}}\>,
\eqno(12) $$
where leptonic masses $m_l$ are $m_e=0.510998918$ MeV for electron and 
$m_\mu=105.658389$ MeV for muon, accordingly, the Mandelstam invariant 
$s=2m_s\mu$ with $\mu=1019.455$ MeV, $N$ is the number of 
colours and $Q=1/3$ for $\phi$-meson. 
To get $<v>$ one may use 
the standard relativistic relation $v=\sqrt{T(T+2E_0)}/(T+E_0)$ with kinetic 
$T$ and rest energies $E_0$ for velocity $v$ of a point-like particle. Putting 
$T=\mu-2m_s$, $E_0=m_s$ we shall gain 
$$<v>=\frac{\sqrt{1-2m_s/\mu}}{1-m_s/\mu} \>.  \eqno(13)$$
At last, obviously, $n=1/V$, while $V$ is the volume of a region where the 
process of annihilation $\bar{s}s \to e^+e^-$ or $\mu^+\mu^-$ occurs, so 
 $V=4\pi<r_{an}>^3/3$ 
with some $<r_{an}>$. 
Then the relations (11)--(13) 
entail the independent estimate for $<r_{an}>$

$$<r_{an}>=\left(\frac{3\sigma\sqrt{2}\sqrt{1-\frac{2m_s}{\mu}}}
{4\pi\Gamma(1-\frac{m_s}{\mu})}\right)^{1/3}           \eqno(14)$$
with $\sigma$ of (12) and $\Gamma=\Gamma_{9,10}$. When inserting $N=3, 
\mu=1019.455$ MeV, $m_s= 107.5$ MeV into (14) we shall have 
$<r_{an}>\approx0.883$ fm for $\Gamma=\Gamma_9$ and 
$<r_{an}>\approx0.889$ fm for $\Gamma=\Gamma_{10}$. 
It should be noted that no experimental values 
exist for the radius $<r>$ in the case of $\phi$-meson\cite{pdg} 
but if taking into account that in the case of proton with 
mass 938 MeV we have $<r>\approx0.875$ fm \cite{pdg}, then if considering 
the size of $\phi$-meson to be approximately the same as for proton we should put, for 
example, $<r>\approx<r_{an}>$ to make computation more sensible. 

In further considerations we can 
use this independent estimate of $<r>$ while calculating $<r>$ according to 
(10) which will impose certain restrictions on parameters of the confining 
SU(3)-gluonic field in $\phi$-meson. 

At last, we should note that there exists the formula of Van Royen--Weisskopf for the 
widths of the leptonic decays of neutral vector mesons (see, e.g., Ref. \cite{Per})

$$\Gamma=\frac{16\pi\alpha_{em}^2Q_V^2}{M_V^2}|\psi(0)|^2 \eqno(14^\prime)$$
with the meson mass $M_V$ and the effective charge $Q_V$ of quarks in the meson so 
$Q_V^2=1/2,1/18,1/9$, respectively, for $\rho$, $\omega$, $\phi$-mesons.

There arises, however, the question: as should be understood $|\psi(0)|^2$ in $(14^\prime)$  
within the framework of our approach because our wave functions (4) consist of three 
components. Probably the most natural prescription 
would be $|\psi(0)|^2=\sum\limits_{j=1}^3|\psi_j(0)|^2$ with $\psi_j$ of (4). But then 
many combinations are possible. E.g., $|\psi_1(0)|\neq 0$, while 
$|\psi_2(0)|=|\psi_3(0)|=0$, or $|\psi_1(0)|=|\psi_2(0)|=0$ while $|\psi_3(0)|\neq 0$ 
and so on. Every such a choice has its physical interpretation and entails its own 
estimates for the gluonic field parameters. The analysis of all the possibilities is 
worth writing the separate paper. Under this situation we decided in the given paper to 
restrict ourselves to a simpler estimate for $<r>$ adduced above in this Section. 

\section{Estimates for parameters of SU(3)-gluonic field in $\phi$-meson}
\subsection{Basic equations}
Now we should consider the equations (8) and (10) as a system which should be 
solved compatibly when $\mu=1019.455$ MeV, $m_s= 107.5$ MeV, 
$g\approx3.771$ while the possible value of $<r>$ has been estimated 
in previous section. While computing 
for distinctness we take all eigenvalues $\lambda_j$ of the Euclidean Dirac 
operator ${\mathcal D}_0$ on the unit 2-sphere ${\mathbb S}^2$ equal to 1.
\subsection{Chiral limit and numerical results}
As was remarked in Refs. \cite{{Gon08b},{Gon10}}, the Dirac equation in the field (1) 
possesses a nontrivial spectrum of bound 
states even for massless fermions. As a result, 
mass of any meson remains nonzero in chiral limit when masses of quarks 
$m_q\to0$ and meson masses will only be expressed through the parameters of 
the confining SU(3)-gluonic field of (1). This purely gluonic residual mass of 
meson should be interpreted as a gluonic contribution to the meson mass.

I.e., the confinement mechanism under consideration gives us  
a possible approach to the problem of chiral symmetry breaking in QCD 
\cite{{Gon08b},{Gon10}}: in chirally symmetric world masses of mesons are fully 
determined by the confining SU(3)-gluonic field between (massless) quarks 
and are not equal to zero. Accordingly chiral symmetry is a sufficiently rough 
approximation holding true only when neglecting the mentioned SU(3)-gluonic 
field between quarks and no additional mechanism of the spontaneous chiral 
symmetry breaking connected to the so-called Goldstone bosons is required. 
Referring for more details to \cite{{Gon08b},{Gon10}}, we can here only say 
that, e.g., mass of $\phi$-meson has also a purely 
gluonic contribution and we may be interested in what part of the $\phi$-meson 
mass is obligatory to that contribution. Indeed, in chiral limit 
$m_{q_1}, m_{q_2}\to0$ 
we obtain from (8)
$$(\mu)_{chiral}\approx\frac{g^2a_1b_1}{\lambda_1-gB_1}\approx
\frac{g^2a_2b_2}{\lambda_2-gB_2}\approx$$
$$\frac{g^2(a_1+a_2)(b_1+b_2)}{\lambda_3+g(B_1+B_2)}\ne0
\>.\eqno(15)$$
and we can see that in 
chiral limit the meson masses are completely determined only 
by the parameters $a_j, b_j, B_j$ of SU(3)-gluonic field between quarks, i.e. 
by interaction between quarks, and those masses have the purely gluonic nature. 
So one can use the parameters $g, a_j, b_j, B_j$ to compute 
$(\mu)_{chiral}$ which in fact represents the sought gluonic 
contribution to the meson masses. The results of the numerical compatible 
solving of equations (8) and (10) are gathered in 
table 1 and 2.

\begin{table}[t]
\caption{Gauge coupling constant, mass parameter $\mu_0$ and
parameters of the confining SU(3)-gluonic field for $\phi$-meson}
\centering
\begin{tabular}{lllllllll}
\hline\noalign{\smallskip}
\small Particle & \small $ g$ & \small $\mu_0$  & \small $a_1$ 
& \small $a_2$ & \small $b_1$  & \small $b_2$ 
& \small $B_1$ & \small $B_2$ \\  
& &(\small MeV) & & &(\small GeV) &(\small GeV) & &  \\  
\tableheadseprule\noalign{\smallskip}
%\hline
%\noalign{\hrule}\\
$\phi$---$\bar{s}s$  & \scriptsize 3.771 & \scriptsize 53.750 & 
\scriptsize 0.668
& \scriptsize -0.263 & \scriptsize 0.223 & \scriptsize -0.536  & 
\scriptsize -0.450
& \scriptsize  -0.440 \\ 
\noalign{\smallskip}\hline
\end{tabular}
\end{table}

\begin{table}[t]
\caption{Theoretical, experimental, chiral $\phi$-meson masses and radius}
\begin{tabular}{lllllllll}
\hline\noalign{\smallskip}
\small Particle & 
\small Theoret. &  
\small Experim.  & 
\small Chiral &
\small Gluonic contribution & 
\small Theoret. $<r>$  & 
\small Experim. $<r>$  \\
& \small (MeV)  
& \small (MeV) 
& \small (MeV)
& (\%)
& \small(fm)  
& \small(fm)\\ 
%\hline
\tableheadseprule\noalign{\smallskip}
\scriptsize $\phi$---$\bar{s}s$   
& \scriptsize $\mu= 2m_s+
\omega_j(0,0,1)= 1019.455$ 
& \scriptsize 1019.46 
& \scriptsize 767.740
& \scriptsize 75.310
& \scriptsize 0.861
& \scriptsize -- \\  
\noalign{\smallskip}\hline
\end{tabular}
\end{table}

\section{Estimates of gluon concentrations, electric and magnetic colour field 
strengths}
\subsection{Estimates}
Now let us recall that, according to Refs. \cite{{Gon052},{Gon04}}, one can 
confront the field (1) with $T_{00}$-component (volumetric energy 
density of the SU(3)-gluonic field) of the energy-momentum tensor (3) so that 
$$T_{00}\equiv T_{tt}=\frac{E^2+H^2}{2}=\frac{1}{2}\left(\frac{a_1^2+
a_1a_2+a_2^2}{r^4}+\frac{b_1^2+b_1b_2+b_2^2}{r^2\sin^2{\vartheta}}\right)
\equiv\frac{{\cal A}}{r^4}+
\frac{{\cal B}}{r^2\sin^2{\vartheta}}\>\eqno(16)$$
with electric $E$ and magnetic $H$ colour field strengths and real 
${\cal A}>0$, ${\cal B}>0$. One can also introduce magnetic colour 
induction $B=(4\pi\times10^{-7} {\rm H/m})\,H$, where $H$ in A/m.

To estimate the gluon concentrations
we can employ (16) and, taking the quantity
$\omega= \Gamma$, the full decay width of a meson, for 
the characteristic frequency of gluons we obtain
the sought characteristic concentration $n$ in the form
$$n=\frac{T_{00}}{\Gamma}\> \eqno(17)$$
so we can rewrite (16) in the form 
$T_{00}=T_{00}^{\rm coul}+T_{00}^{\rm lin}$ conforming to the contributions 
from the Coulomb and linear parts of the
solution (1). This entails the corresponding split of $n$ from (17) as 
$n=n_{\rm coul} + n_{\rm lin}$. 

The parameters of Table 1 were employed when computing and for simplicity 
we put $\sin{\vartheta}=1$ in (16). Also there was used the following 
present-day full decay width of $\phi$-meson ${\Gamma}=4.26$ MeV, 
whereas the Bohr radius 
$a_0=0.529\cdot10^{5}\ {\rm fm}$ \cite{pdg}.   

Table 3 contains the numerical results for $n_{\rm coul}$, $n_{\rm lin}$, $n$, 
$E$, $H$, $B$ for the meson under discussion.

\begin{table}[t]
\caption{Gluon concentrations, electric and magnetic colour field strengths in 
$\phi$-meson}
\begin{tabular}{lllllll}  
\hline\noalign{\smallskip}
\scriptsize $\phi$---$\bar{s}s$: 
& \scriptsize $r_0=<r>= 0.861 \ {\rm fm}$ 
& 
& 
&  
& 
&\\
\hline
\tiny $r$ 
& \tiny $n_{\rm coul}$ 
& \tiny $n_{\rm lin}$ 
& \tiny $n$ 
& \tiny $E$ 
& \tiny $H$ 
& \tiny $B$\\
\tiny (fm) 
& \tiny $ ({\rm m}^{-3}) $ 
& \tiny (${\rm m}^{-3}) $ 
& \tiny (${\rm m}^{-3}) $ 
& \tiny $({\rm V/m})$ 
& \tiny $({\rm A/m})$
& \tiny $({\rm T})$ \\ 
\tableheadseprule\noalign{\smallskip}
\tiny $0.1r_0$ 
& \tiny $ 0.942\times10^{53}$   
& \tiny $ 0.448\times10^{51}$ 
& \tiny $ 0.947\times10^{53}$ 
& \tiny $ 0.132\times10^{25}$  
& \tiny $ 0.123\times10^{22}$ 
& \tiny $ 0.154\times10^{16}$\\
%\hline
\tiny$r_0$ 
& \tiny$ 0.942\times10^{49}$ 
& \tiny$ 0.448\times10^{49}$ 
& \tiny$ 0.139\times10^{50}$
& \tiny$ 0.132\times10^{23}$  
& \tiny$ 0.123\times10^{21}$  
& \tiny$ 0.154\times10^{15}$ \\
%\hline
\tiny$1.0$ 
& \tiny$ 0.518\times10^{49}$  
& \tiny$ 0.332\times10^{49}$ 
& \tiny$ 0.850\times10^{49}$ 
& \tiny$ 0.982\times10^{22}$  
& \tiny$ 0.106\times10^{21}$  
& \tiny$ 0.133\times10^{15}$ \\
%\hline
\tiny$10r_0$ 
& \tiny$ 0.942\times10^{45}$  
& \tiny$ 0.448\times10^{47}$ 
& \tiny$ 0.458\times10^{47}$ 
& \tiny$ 0.132\times10^{21}$  
& \tiny$ 0.123\times10^{20}$  
& \tiny$ 0.154\times10^{14}$ \\
%\hline
\tiny$a_0$ 
& \tiny$ 0.661\times10^{30}$  
& \tiny$ 0.119\times10^{40}$ 
& \tiny$ 0.119\times10^{40}$ 
& \tiny$ 0.351\times10^{13}$ 
& \tiny$ 0.200\times10^{16}$  
& \tiny$ 0.251\times10^{10}$ \\ 
\noalign{\smallskip}\hline
\end{tabular}
\end{table}

\subsection{Concluding Remarks}

 As is seen from Table 3, at the characteristic scales
of $\phi$-meson the gluon concentrations are huge and the corresponding 
fields (electric and magnetic colour ones) can be considered to be 
the classical ones with enormous strenghts. The part $n_{\rm coul}$ of gluon 
concentration $n$ connected with the Coulomb electric colour field is 
decreasing faster than $n_{\rm lin}$, the part of $n$ related to the linear 
magnetic colour field, and at large distances $n_{\rm lin}$ becomes dominant. 
It should be emphasized that in fact the gluon concentrations are much 
greater than the estimates given in Table 3 
because the latter are the estimates for maximal possible gluon frequencies, 
i.e. for maximal possible gluon impulses (under the concrete situation of 
$\phi$-meson). The given picture is in concordance with the one obtained 
in Refs. \cite{{Gon07a},{Gon07b},{Gon08},{Gon08b},{Gon10}}. As a result, the confinement 
mechanism developed in 
Refs. \cite{{Gon01},{Gon051},{Gon052}} is also confirmed by the considerations 
of the present paper. 

It should be noted, however, that our results are of a preliminary character 
which is readily apparent, for example, from that the current quark masses 
(as well as the gauge coupling constant $g$) used in computation are known only within the 
certain limits and we can expect similar limits for the magnitudes 
discussed in the paper so it is neccesary further specification of the 
parameters for the confining SU(3)-gluonic field 
in $\phi$-meson which can be obtained, for instance, by calculating 
widths of radiative decays of type $\phi\to\pi^0+\gamma$, $\phi\to\eta+\gamma$  
and so on \cite{pdg}. We hope to continue analysing the given problems 
elsewhere. 

Finally, one can emphasize that though there exists a number of papers devoted 
to miscellaneous aspects of the vector meson physics (see e.g. 
Refs. \cite{{Lut},{Gia},{Aoki}} and references therein) but all of them do not 
directly appeal to the quark and gluonic degrees of freedom as should be from 
the first principles of QCD. Really, they use the so-called potential approach where 
interaction between quarks is described by potential of form $V(r)=a/r+br$ with some 
constants $a$ and $b$. We cannot, however, speak about 
the above potential $V(r)$ as describing some gluon configuration between quarks. 
It would be possible if the mentioned potential were a solution of Yang-Mills 
equations directly derived from QCD-Lagrangian since, from the QCD-point of 
view, any gluonic field should be a solution of Yang-Mills equations (as well 
as any electromagnetic field is by definition always a solution of Maxwell 
equations). But in virtue of the uniqueness theorem of Section 1  
it is impossible: Coulomb and linear parts belong to different parts of the 
YM-potentials, accordingly, to the colour electric and colour magnetic parts 
so they cannot be united in one component of form $a/r+br$. 

On the contrary, within the framework of our 
approach the words {\em quark and gluonic degrees of freedom} make exact sense: 
gluons come forward in the form of bosonic condensate described by parameters 
$a_j$, $b_j$, $B_j$ from the unique exact solution (1) of the Yang-Mills 
equations while quarks are represented by their current masses $m_q$.

\section{Conclusion}
The main idea of quark confinement may be borrowed from classical electrodynamics.
Indeed,  let us recall the well-known case of
motion of a charged particle in the homogeneous magnetic field
(see, e.g., Ref. \cite{LL}). In the latter case the particle moves along helical curve
with lead of helix $h=2\pi mv\cos{\alpha}/(qH\sqrt{1-v^2})$ and radius
$R=mv\sin{\alpha}/(qB\sqrt{1-v^2})$, where $\alpha$ is an angle between vectors of
the particle velocity $\bf v$ and magnetic induction $\bf B$, $q$ is a particle charge,
$m$ is a particle mass. As a consequence, the homogeneous magnetic
field does not give rise to the full confinement of the particle since the
latter may go to infinity along the helical curve. The situation is not changed
at quantum level as well: there exist no bound states in the homogeneous
magnetic field \cite{LLQ}. But if estimating module of $B$ at
$R\sim10^{-15}\,{\rm m}=1$ fm for electron then $B$ will be of order $10^{23}$ T. I. e.,
if considering that quarks are confined by a colour magnetic field that they themselves
create then one needs a colour magnetic field between quarks with $B$ of such an order
and that field should not allow any quark to go to infinity. It is clear this field should
be a solution of the Yang-Mills equations. Really, if speaking about Minkowski spacetime
then searching for classical solutions of the Yang-Mills equations makes sense because
at large distances between quarks the latter are surrounded with a huge number of gluons
that are emitted by both quarks and gluons themselves. Under this situation it is quite
plausible that confinement of quarks arises due to certain properties of such gluonic
clouds while the latter should be described just by {\em classical} solutions of
the Yang-Mills equations.

As was shown in Refs. \cite{{Gon01},{Gon051},{Gon052}} (see also Appendix C in 
Ref. \cite{Gon10}), the necessary solution has the
form (1) for group SU(3) and is unique in a certain sense.
This result allowed us to propose a quark confinement mechanism which was successfully
applied to meson spectroscopy. Indeed, as follows from (10) at $|b_j|\to\infty$ we
have $<r>\,\sim\, \sqrt{\sum\limits_{j=1}^3\frac{1}
{(g|b_j|)^2}}$, so in the strong magnetic colour field when $|b_j|\to\infty$,
$<r>\to 0$, while the meson wave functions of (4) and (6) behave as
$\Psi_j\,\sim\,e^{-g|b_j|r}$, i. e., just the magnetic colour field of (1)
provides two quarks with confinement. This situation also holds true at classical
level \cite{GF10}.

The given paper extends this mechanism over vector mesons and further 
study of that approach will be connected with analysis of other concrete mesons and 
baryons with its help.

%\begin{acknowledgements}
%If you'd like to thank anyone, place your comments here
%and remove the percent signs.
%\end{acknowledgements}

% BibTeX users please use
%\bibliographystyle{spbasic}
%\bibliography{}   % name your BibTeX data base

% Non-BibTeX users please use

\end{document}